\documentclass[journal]{vgtc}

\usepackage[english]{babel}
\usepackage[utf8x]{inputenc}
\usepackage[T1]{fontenc}
\usepackage{gensymb} 

\usepackage{dsfont}

\usepackage{amsmath}
\usepackage{graphicx}
\usepackage[colorinlistoftodos]{todonotes}
\usepackage[colorlinks=true, allcolors=blue]{hyperref}

\usepackage{times}

\begin{document}

\title{Why scatter plots suggest causality, and what we can do about it.}

\author{Carl T. Bergstrom, Jevin D. West}

\authorfooter{
\item
 Carl T. Bergstrom, Department of Biology, University of Washington: cbergst@uw.edu.
\item
 Jevin D. West, Information School, University of Washington: jevinw@uw.edu.
\item Date submitted: September 23, 2018
}



\abstract{Scatter plots carry an implicit if subtle message about causality. Whether we look at functions of one variable in pure mathematics, plots of experimental measurements as a function of the experimental conditions, or scatter plots of predictor and response variables, the value plotted on the vertical axis is by convention assumed to be determined or influenced by the value on the horizontal axis. This is a problem for the public understanding of scientific results and perhaps also for professional scientists' interpretations of scatter plots. To avoid suggesting a causal relationship between the $x$ and $y$ values in a scatter plot, we propose a new type of data visualization, the diamond plot. Diamond plots are essentially 45$\degree$ rotations of ordinary scatter plots; by visually jarring the viewer they clearly indicate that she should not draw the usual distinction between independent/predictor variable and dependent/response variable. Instead, she should see the relationship as purely correlative.}



\maketitle

\section{Introduction: Correlation and Causation}

\bigskip

\noindent {\em ``Correlation does not imply causation."}

\bigskip

Many students are taught this phrase in high school, or if not there, in the course of a college curriculum in any STEM (science, technology, engineering, and medicine) discipline. Of those students, a reasonable fraction seem to understand the basic point. But surprisingly few internalize the full importance of this simple phrase. 

As a result, when laypeople---and many professional scientists as well---are presented with evidence of a correlation, they often draw unmerited inferences about causation. This is not solely their fault; those presenting data often nudge readers toward these unfounded causal conclusions. Reporters conflate correlation and causation in the popular news media; press officers describe correlations in causal terms in press releases; researchers make causal claims based on evidence of correlative evidence in the scientific literature.  

We suspect that these mistakes most often arise through a confluence of misunderstanding and wishful thinking. After all, causal results are considerably more interesting, valuable, and even exciting than correlations.  Correlations hint at how the world might work, causal relationships directly answer the question. Correlations might be spurious; casual relationships are the real thing. Perhaps most importantly, correlations don't tell us what to do to change a system in the direction we want, whereas causal relationships allow us to suggest possible interventions to bring about the outcomes we desire.

Every week brings a host of new examples. A {\em New York Times} article described an observational study that demonstrated an association between playing tennis and increased lifespan \cite{trytennis}. The abstract of the study makes it clear that there is no implication of causation: "Because this is an observational study, it remains uncertain whether this relationship is causal." Yet the title of the {\em New York Times} story is blatantly prescriptive: ``The Best Sport for a Longer Life? Try Tennis". 

A research team from real estate website Zillow.com published a white paper about a negative correlation between changes in housing prices and changes in fertility rates \cite{zillow}. The study carefully disclaimed any causal implications: 

\begin{quotation}
As a further caveat, the correlation observed here is by no means proof that home value growth causes fertility declines. One alternative explanation could be the possibility that there is clustering into certain counties of people with careers that pay well enough for expensive homes but make it difficult to have children before 30; this could cause both trends observed in the chart above. There are many other confounding factors that could explain this relationship as well, such as the possibility that cultural values or the cost of child care varies across counties with some correlation to home values.
\end{quotation}

This is an example of how correlations should be presented to help readers avoid assuming causality.  But apparently it wasn't enough for the financial information website {\em Marketwatch} \cite{marketwatch}. The title of their story about the Zillow team's findings implies a causal relationship: ``Another adverse effect of high home prices: Fewer babies". The first line of the story does the same: “Forget about a baby boom — rising home prices appear to be causing many would-be parents to think twice before expanding their family.” All this, despite the inclusion later in the {\em Marketwatch} of a paragraph explaining that the study does not determine the direction of causality. 

Journalists and the news media are not alone in shouldering the blame. Researchers make the same mistakes. A recent scientific paper reported an observational study of milk fat consumption and childhood obesity \cite{beck2017full}. Despite explicitly acknowledging the inability to draw causal inferences, the paper's title implies causality:  "Full fat milk consumption protects against severe childhood obesity in Latinos". Worse yet, the authors use their findings to question prescribing practices: "These results call into question recommendations that promote consumption of lower fat milk." 

In short, correlations are commonly misinterpreted as evidence of causation. The problem is exacerbated when authors and reporters, under pressure to provide novel and meaningful findings, shade in this direction to make their results seem more exciting and explanatory. 

\section{Scatter plots and why they can suggest causality}

The scatter plot---also known as the scatter graph or scatter diagram---is among the most common forms of data visualization in the scientific literature. This staple of data display is used to represent bivariate data $(x_i,y_i)$, with the intent of revealing relationships between the $x$ and $y$ values of each pair $i$ \cite{anscombe,friendly2005early,sarikaya2018scatterplots}. (The major alternative for displaying paired data of this type is the "parallel coordinates plot" \cite{kanjanabose2015multi}; this type of visualization can be effective for small data sets with high correlations, but rapidly becomes difficult to interpret for larger data sets and for lower degrees of correlation.)

Thirty years ago, data graphics that compared multiple variables were extremely rare in the popular media (with the exception of time-courses) \cite{tuftevisual}. No longer. The scatter plot has made its way from the pages of scholarly articles into the public sphere. By and large this is a positive development, in that it means that the public are engaging with data at a deeper level than a few decades ago. But with this greater sophistication comes the risk of misunderstanding and misinterpretation particularly---though not exclusively---among those without formal training. In this note, we focus on one particular risk, the subtle and often unintended implication of a causal relationship that comes with a scatter plot.

The fundamental problem with scatter plots is that many aspects of STEM education use the Cartesian plane to represent cause-and-effect or dependency relationships. In many plots on the Cartesian plane, the $x$ value on the horizontal axis has some role in determining or influencing the $y$ value on the vertical axis. A high school or college student is likely to internalize these conventions; when viewing a plot in the Cartesian plane the natural assumption is that the variable on the horizontal axis is having some causal influence over the value on the vertical axis. 

\begin{figure}
\includegraphics[width=.48 \textwidth]{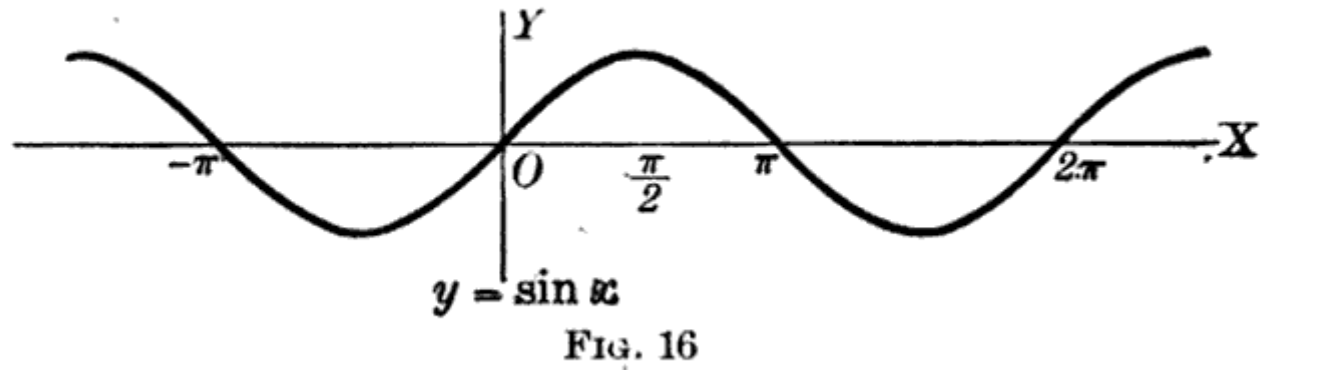}
\caption{ For this mathematical function, the value along the vertical axis, $y$ is explicitly determined by the value $x$ along the horizontal axis. Shown here, a graph of the function $y=\sin{x}$ from a 1916 calculus textbook \cite{love1916differential}.}
\label{fig:fofx}
\end{figure}

In the study of analytic geometry, trigonometry, and calculus, the Cartesian plane is used to represent a function $y=f(x)$ by representing $x$ on the horizontal axis and $f(x)$ on the vertical axis. Here the $x$ value literally determines the $y$ value by means of the function $f$ (Figure \ref{fig:fofx}). In this application, we are explicitly cautioned against anything short of a purely deterministic relationship between $x$ and $y$: a proper function can take on only one value of $y$ for a given $x$. 

\begin{figure}
\includegraphics[width=.48 \textwidth]{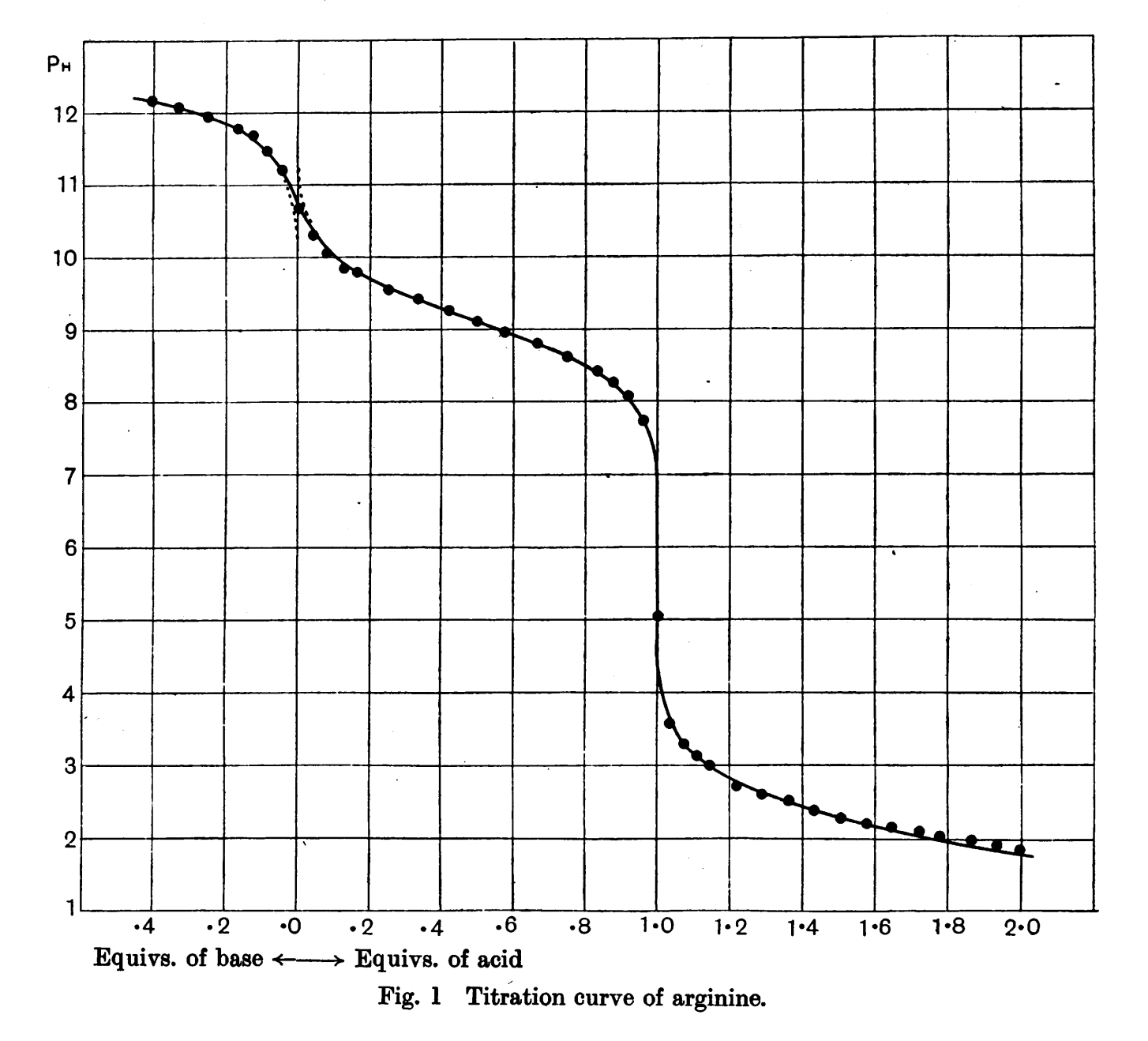}
\caption{In a titration curve, the experimenter controls the volume of titrant added (horizontal axis) and measures the resulting pH of the solution (vertical axis). Shown: a 1924 titration of argenine monohydrochloride with barium hydroxide and hydrochloric acid \cite{hunter1924dissociation}.  }
\label{fig:titration}
\end{figure}

When we learn about experimental methodology in a chemistry or physics lab, we think in terms of independent and dependent variables. Consider a titration curve (Figure \ref{fig:titration}). The independent variable---the volume of titrant added---is shown on the horizontal axis. The dependent variable, the pH of the solution, is shown on the vertical axis. Our aim in this case is precisely to reveal a causal relationship: we want to determine how the titration volume influences pH\footnote{There is an interesting subplot involving how we think about error and variation more broadly. Often when we look at a scatter plot the first inclination is to perform---in practice or simply in one's head---a regression on the data. But standard regression methods assume that the $x$ values are exact and known, where is the $y$ values are subject to error. Hence least squares regression maximizes the sum of squared vertical distances between the regression line and data points. For many scatter plots, both $x$ and $y$ values are subject to measurement error and other sources of noise. So-called Deming regression addresses this issue by minimizing a weighted function of the total squared distance from the regression line. It is not clear to what degree someone viewing a scatter chart make assumptions about the fixity of the $x$ values and the error in $y$ values, but this merits consideration.}.  

\begin{figure}
\includegraphics[width=.48 \textwidth]{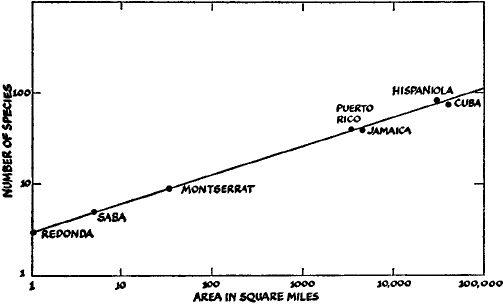}
\caption{One of MacArthur and Wilson's species-area plots from their classic 1967 book {\em The Theory of Island Biography}. This plot shows the number of reptile and amphibian species for islands of the West Indies \cite{macarthur2001theory}. }
\label{fig:MW}
\end{figure}

When looking statistically at observational data in the natural and social sciences, we often distinguish between predictor and response variables. When doing so, we have in mind some underlying model of a causal relationship. The predictor variable causes — or at least influences — the response variable, whereas the converse is not true. Once again the standard convention is to plot the predictor variable on the horizontal axis and the response variable on the vertical axis. For example, MacArthur and Wilson's theory of island biogeography \cite{macarthur2001theory} proposes that the land area of an island determines the number of species that it can support. To support their hypothesis, MacArthur and Wilson presented a number of {\em species-area plots}. These are log-log scatter plots, with each point corresponding to one island, of the number of species on the island plotted as a function of the island's area (Figure \ref{fig:MW}). Note the causal relationship in which the predictor variable, area, has a causal impact on the response variable, species number. Clearly causality cannot be going in the opposite direction. Indeed, MacArthur and Wilson present a theoretical model of how land area causally influences species number.

These conventions are long-standing within the technical literature and have been thoroughly integrated into math and science education at the secondary and post-secondary levels. As a consequence, problems arise on two levels. First, technical readers---even those with the training to know better---may be nudged toward sloppy thinking about causality when data are presented as scatter plots. Second, when scatter plots appear in the news media or trade press, non-technical readers all too often interpret correlative data as evidence of a causal relationship.

\section{What we can do about it}

In a perfect world we might choose to reconstruct the grammar of data visualization from scratch, taking pains to ensure that the Cartesian plane carries no implication of causality. But this is not a perfect world. We are stuck with with the norms we already have.

Given the conventional understanding around the Cartesian plane and around scatter plots in the Cartesian plane in particular, we propose a simple visual cue to indicate that a scatter plot carries no implication of causality. The cue that we suggest is simple: a 45$\degree$ counter-clockwise rotation of the axes, to form a diamond-shaped graph with the origin at the bottom point. While we have not yet run a rigorous observational study, in our experience statistically trained readers initially find diamond plots to be jarring. This is not necessarily a bad thing, given the prevalence of causal misinterpretation. The unexpected form of data visualization draws attention to the issue of causality and makes it clear that no causal relationship is implied.

\begin{figure}
\includegraphics[width=.48 \textwidth]{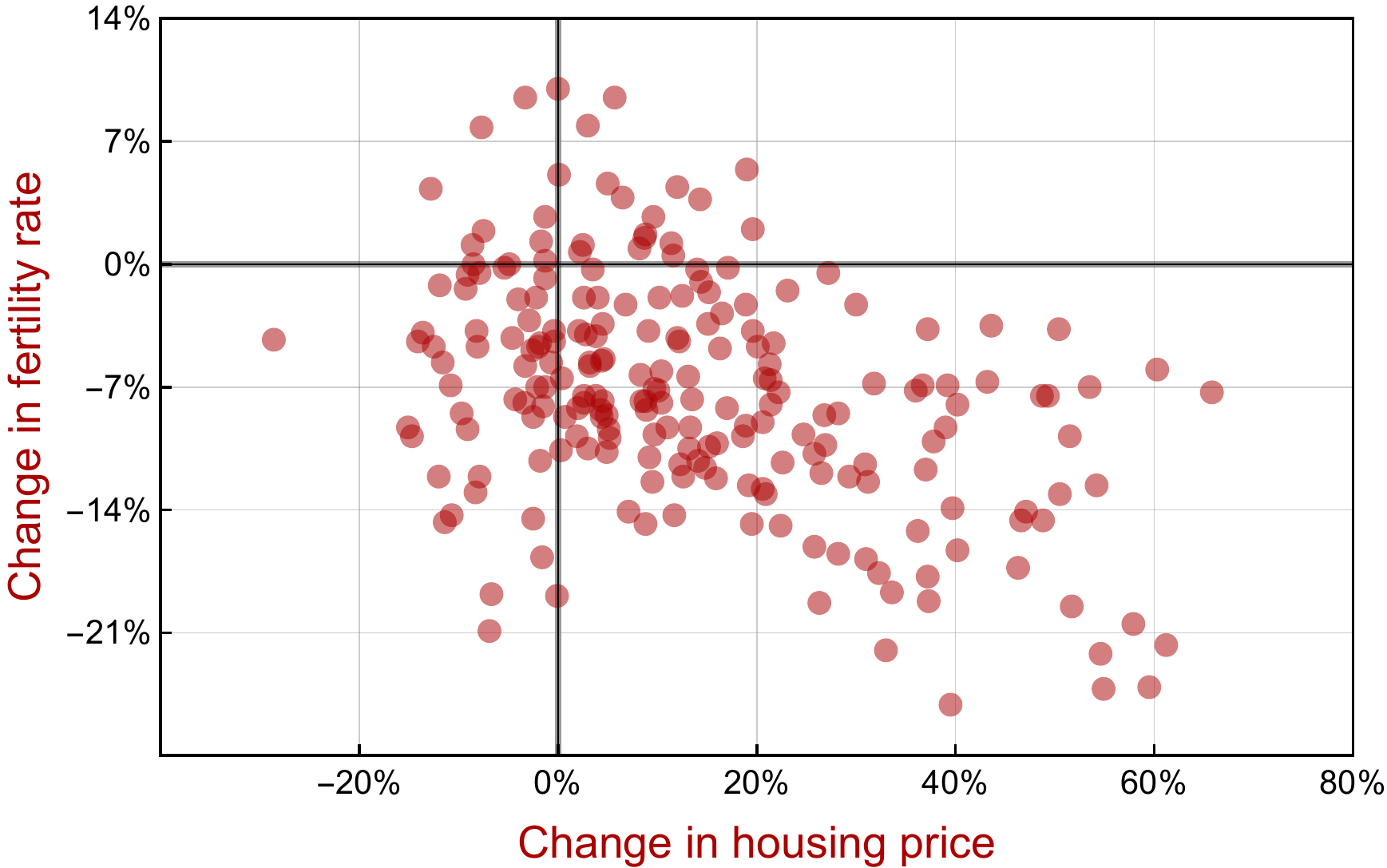}
\caption{A standard scatter plot of the Zillow study described above \cite{zillow}. The horizontal axis represents the percent change in housing value from 2010 to 2016 for counties across the U.S. The vertical axis represents the percent change in fertility rate for 25 to 29 year-old women from 2010 to 2016 for those same counties. }
\label{fig:trad}
\end{figure}

We illustrate this approach using data from the Zillow research study that we described in the introduction. Figure \ref{fig:trad} shows the Zillow data in the form of a traditional scatter plot\footnote{The Zillow team's white paper includes a well-designed interactive data visualization in the form of a bubble chart: a scatter plot where color and point size are used to code two additional variables. For simplicity, we will not use color and point size here, though one could in principle do both of those things in a diamond plot.}. While this graph clearly reveals the correlation between change in housing price and change in fertility rate, it also subtly suggests causality. The choice of horizontal and vertical axes may lead readers to infer that housing prices are driving fertility rates. 

\begin{figure}
\includegraphics[width=.52 \textwidth]{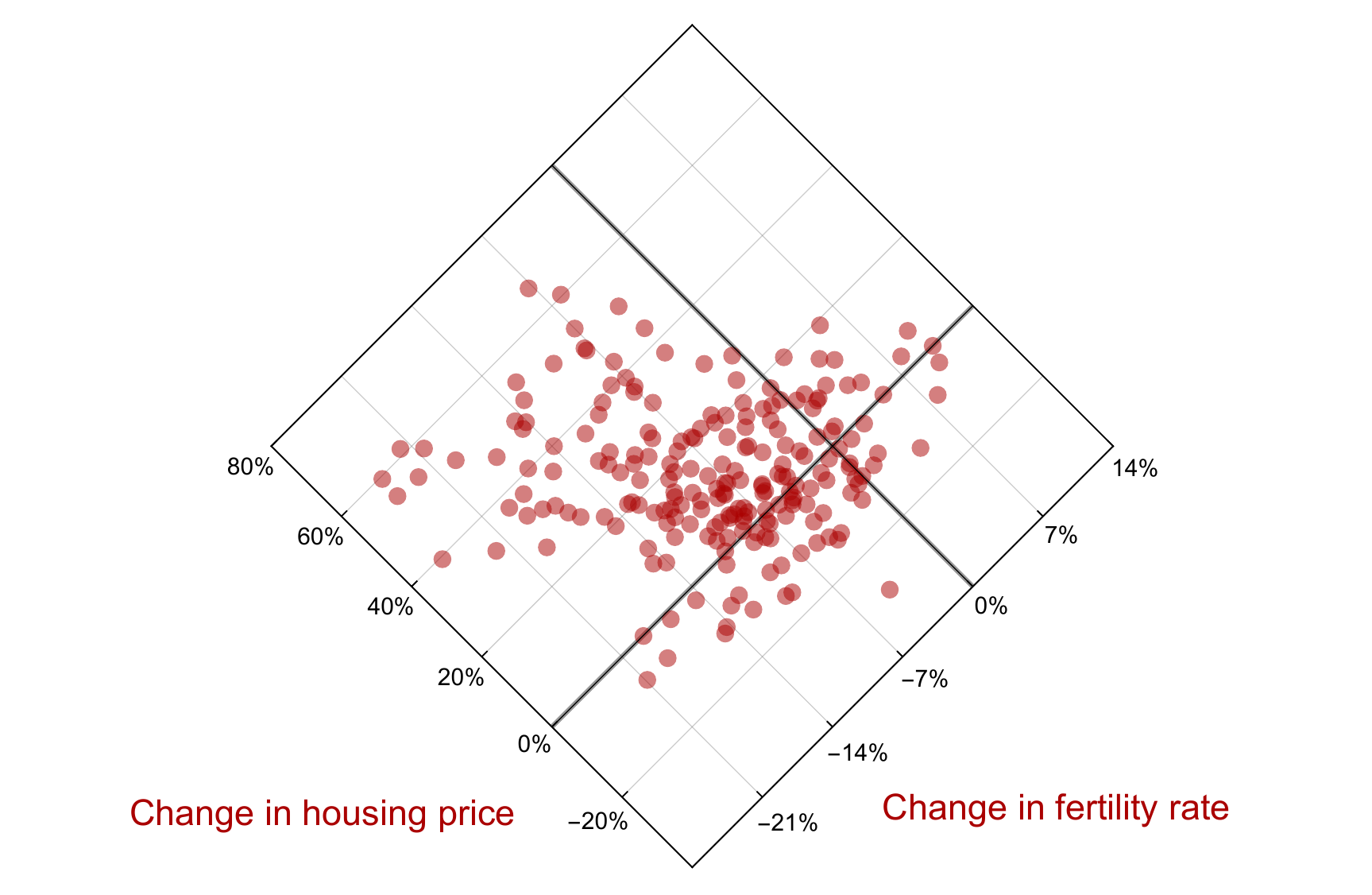}
\caption{A diamond plot of the Zillow data plotted in Figure~\ref{fig:trad}. The {\em diamond plot}  rotates the scatter plot 45$\degree$, adjusts the aspect ratio to 1:1, and realigns the axes titles and labels horizontally. The data graphics for this figure (and figures 4 though 8) were created in {\em Mathematica}. Mathematica code is available at http://www.eigenfactor.org/diamondplot.nb }
\label{fig:zillow_tilted}
\end{figure}

In Figure \ref{fig:zillow_tilted}, we display the same data using a {\em diamond plot} visualization.  Because we have rotated the axes by 45 degrees, neither axis has priority as the dependent or independent variable, or as the response or predictor variable. To a reader familiar with scatter plots, this diamond plot also appears a bit out of the ordinary. By jarring the viewer in this way, it calls attention to the issue of the axes and their causal relationships to one another. It reminds the reader to think about this causality issue explicitly and to avoid unintentionally drawing causal inferences. 

\begin{figure}
    \centering
 \includegraphics[width=.48 \textwidth]{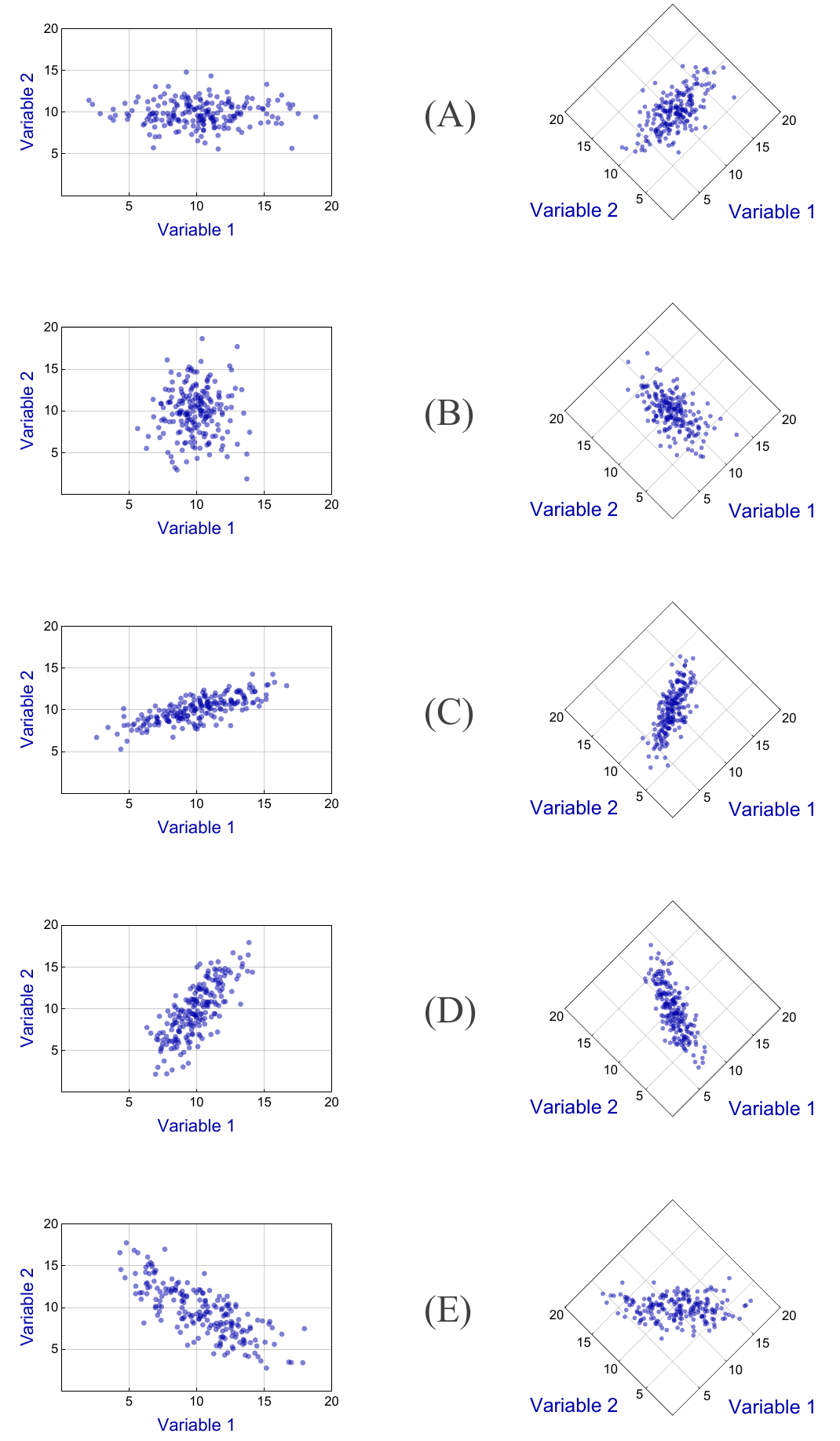}
    \caption{Paired scatter plots and diamond plots of bivariate normal distributions. (A) Variables 1 and 2 are uncorrelated, variable 1 has higher variance. (B) Variables 1 and 2 are uncorrelated, variable 2 has higher variance. (C) Variables 1 and 2 are positively correlated, $\rho=0.75$, variable 1 has higher variance. (D) Variables 1 and 2 are positively correlated, $\rho=0.75$, variable 2 has higher variance. (E) Variables 1 and 2 are negatively correlated, $\rho=-0.75$ with equal variances.}
    \label{fig:examples}
\end{figure}

\begin{figure}
    \centering
 \includegraphics[width=.48 \textwidth]{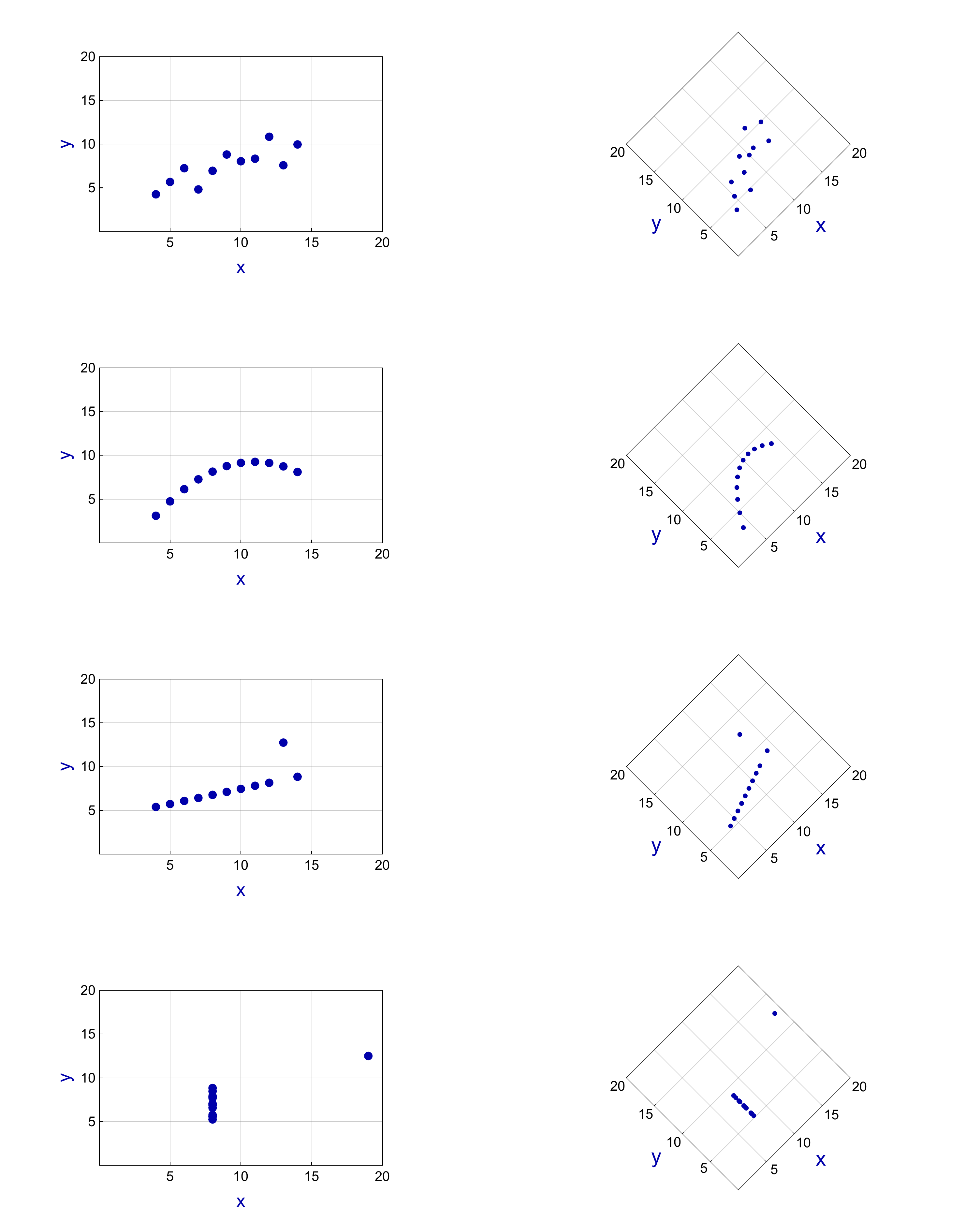}
    \caption{Paired scatter plots and diamond plots for Anscombe's quartet. The data sets all have the same mean, variance and correlation values but are quite different when presented visually. }
    \label{fig:anscombe}
\end{figure}

On a diamond plot, uncorrelated data will take the form of an oval with a diagonal axis, whereas correlated data will deviate from this diagonal. In figure \ref{fig:examples}, we show scatter plots and their corresponding diamond plots for a series of bivariate normal distributions. When the two variables are uncorrelated, the distributions appear along one of the two 45 degree diagonals in the diamond plot \ref{fig:examples}A,B. When the variables are correlated, the shape of the distribution shifts away from the diagonal toward the vertical (for positive correlations; \ref{fig:examples}C,D) or horizontal (for negative correlations; \ref{fig:examples}E). In figure \ref{fig:anscombe}, we show Anscombe's famous quartet \cite{anscombe} of bivariate distributions,  each in its original scatter plot form, and then as a diamond plot. Anscombe's quartet features four different scatter plots, each with the same mean and variance for both variables, the same linear fit, and the same correlation coefficient (0.816). Anscombe intended these distributions to illustrate how descriptive statistics can be inadequate if not outright misleading --- and how data visualizations can better differentiate these patterns. We show them here as illustrations of how data appear on a diamond plot, because collectively they span many of the patterns one might see underlying a positive correlation coefficient.  

Several considerations influenced our design decision to rotate the axes 45$\degree$ counter-clockwise but maintain horizontal axis titles. First, the graphic has to be similar enough to a standard scatter plot that users will recognize the form of data presentation. At the same time, it has to be different enough that the user would notice something distinct and important about the data presentation --- namely, that the bivariate data presented carry no explicit causal relationship. Second, the graphic needs to treat the two axes as symmetrically as possible, so that neither is prioritized over the other as an independent variable. These considerations led us to (1) the 45$\degree$ rotation of a conventional scatter plot, and (2) the adoption of a 1:1 aspect ratio instead of the 1.61:1 golden ratio that we favor for scatter plots.

Additional design decisions pertained to the details of the rotation. We wanted to provide a rotation that required the  least cognitive effort in reorientation. Turning counterclockwise places the origin---often (0,0)---at the lowest point on the graph and ensures that both axes increase in value as they move up the page. A 45$\degree$ rotation preserves the desired symmetry between the two axes; one would be privileged over the other if the axes were rotated otherwise, e.g. 30$\degree$ or 60$\degree$. While we feel generally feel that grid lines are to be used with considerable trepidation and a light touch, we find them essential in diamond plots because of the unusual 45$\degree$ rotation of the axes.   Another design concern involves how to orient the axis titles and labels. Again, our thoughts were that the titles should be easy to read and interpret with this new orientation. Thus we decided to keep the axes titles horizontal, to reduce the effort in reading the plot. Doing so also creates a pleasant `grounding' effect in which the tilted diamond is visually stabilized by the platform that the axis labels create.

\begin{figure*}
\includegraphics[width=\textwidth]{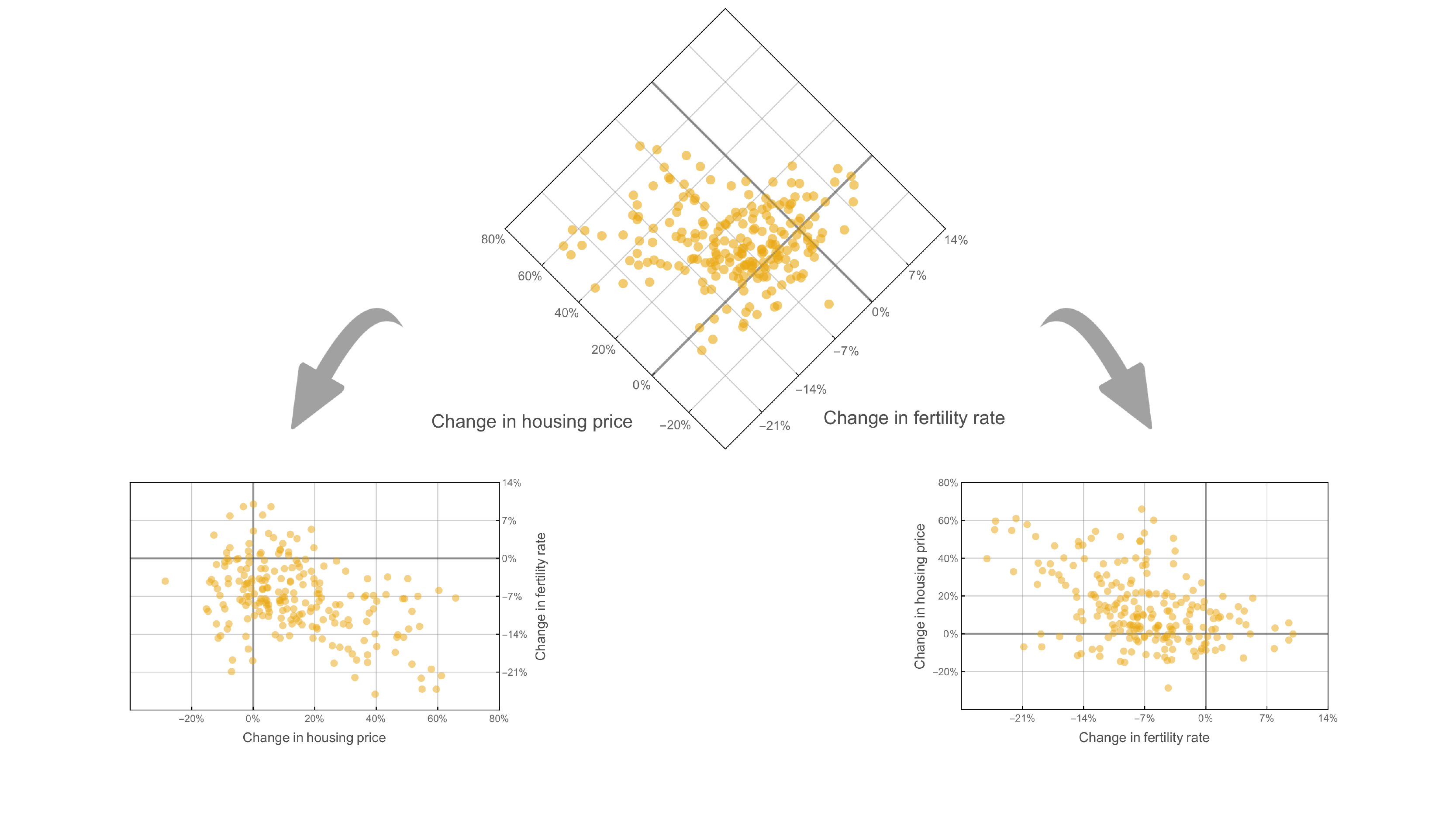}
\caption{The design for an interactive diamond plot. The interactive version would allow the user to rotate to view the conventional scatter plots with either assignment of variables to the horizontal and vertical. The equal emphasis on both directions of rotation highlights the fact that neither variable is privileged as an independent variable.}
\label{fig:interactive}
\end{figure*}

Whether in the popular media or in scientific manuscripts, data graphics are increasingly viewed in an online environment. This makes interactive visualization possible. Diamond plots are readily extended to an interactive form that allows the viewer to see the data in more familiar scatter plot contexts without any suggestion of causality. In figure \ref{fig:interactive} we sketch out the basic design. The figure first appears as a diamond plot, but the user can drag clockwise or counterclockwise, rotating the graph 45 degrees in either direction to construct either of the two possible scatter plots: one with variable 1 on the horizontal and variable 2 on the vertical, and one with the converse arrangement. Notice that the counterclockwise rotation also requires a ``flip'', reversing the vertical axis to place the origin at the lower left instead of lower right corner. While unfortunate, this is inevitable because inter-converting between the two scatter plots requires a flip as well as a rotation.

In order to evaluate our design decisions and assess the merits and disadvantages of diamond plots relative to scatter plots, we will need to conduct a user study to evaluate whether, at minimum, the following are true. (1) Traditional scatter plots nudge viewers to assume causality when no such assumption is merited, and (2) Diamond plots ameliorate this problem. In addition, we need (3) to assess the additional cognitive load required to interpret data in diamond plot form. Are there time and accuracy costs associated with performing standard tasks---assessing correlation direction and magnitude, comparing variances, identifying outliers---using diamond plots rather than scatter plots. We are planning such a study, but our initial feedback, through conversations with colleagues and users on twitter, has been generally quite positive. 

\section{Conclusion}

Misinterpreting scatter plots --- by assuming the horizontal axis variable causally influences the  vertical axis variable --- is a common mistake in science, journalism and business. Such mistakes can lead to erroneous causal conclusions and unfounded prescriptive recommendations. To address this, we propose a variant on the scatter plot, a 45$\degree$ rotation that we call a diamond plot. Our hope is that this subtle design modification will reduce these misinterpretations and assist with accurate causal reasoning about data.  

\section*{Acknowledgments}

We thank the Twitter community for showing sufficient enthusiasm around these issues to motivate the present note. In the course of that Twitter discussion (https://twitter.com/CT\_Bergstrom/status/1035327464644333568), Detlef Weigel (@plantlab) and Taylor Kessinger (@koaleszenz) validated our design intuitions by independently suggesting the same 45$\degree$ rotation we were considering. 
We thank Holly Bergstrom for the encouragement we needed to actually write all of this down.


\end{document}